\documentclass[12pt,preprint]{aastex}

\def\xmm{{\it XMM-Newton}}
\def\planck{{\it Planck}}
\def\plck{PLCK~G287.0+32.9\,}
\def\lesssim{\mathrel{\hbox{\rlap{\hbox{\lower4pt\hbox{$\sim$}}}\hbox{$<$}}}}
\def\gtrsim{\mathrel{\hbox{\rlap{\hbox{\lower4pt\hbox{$\sim$}}}\hbox{$>$}}}}
\newcommand{\lapp}{\mbox{\raisebox{-0.3em}{$\stackrel{\textstyle <}{\sim}$}}}

\def\arcsec{$^{\prime\prime}\,$}
\def\arcmin{$^{\prime}\,$}
\def\degrees{$^{\circ}$\,}

\shorttitle{{\em Planck} ESZ cluster PLCK~G287.0+32.9: diffuse non-thermal radio emission}
\shortauthors{Bagchi et al.}

\begin{document}

\title{DISCOVERY OF THE FIRST GIANT DOUBLE RADIO RELIC
IN A GALAXY CLUSTER FOUND IN THE {\it PLANCK}
SUNYAEV-ZEL'DOVICH CLUSTER SURVEY: PLCK~G287.0+32.9
}

\author{Joydeep Bagchi\altaffilmark{1}, S. K. Sirothia\altaffilmark{2}, 
Norbert Werner\altaffilmark{3}, Mahadev B. Pandge\altaffilmark{4},
     Nimisha G. Kantharia\altaffilmark{2}, C.H. Ishwara-Chandra\altaffilmark{2}, Gopal-Krishna\altaffilmark{2}, 
     Surajit Paul\altaffilmark{1} and   Santosh Joshi\altaffilmark{5}}
\altaffiltext{1}{The Inter-University Centre for Astronomy and Astrophysics (IUCAA),  Pune University Campus, Post Bag 4, Pune  411007, India; joydeep@iucaa.ernet.in}
\altaffiltext{2}{National Centre for Radio Astrophysics, TIFR, Pune University
Campus, Post Bag 3, Pune  411007, India}
\altaffiltext{3}{Kavli Institute for Particle Astrophysics and Cosmology, Stanford
University, 452 Lomita Mall, Stanford, CA 94305, USA}
\altaffiltext{4}{School of Physical Sciences, Swami Ramanand Teerth Marathwada University, Vishnupuri, Nanded 431606, India}
\altaffiltext{5}{Aryabhatta Research Institute of Observational Sciences (ARIES), Manora Peak,
Nainital 263129, India}


\begin{abstract}
We report the discovery of large-scale diffuse non-thermal radio emission in  
PLCK~G287.0+32.9, an exceptionally hot ($T\sim $ 13 keV), massive, and luminous 
galaxy cluster, strongly detected by the \planck\ satellite in a recent, all-sky blind search for new 
clusters through Sunyaev--Zel'dovich effect. 
Giant Metrewave Radio
Telescope 150 MHz and Very Large Array 1.4 GHz radio data reveal  a pair of giant ($>$1 Mpc) 
``arc''  shaped peripheral radio-relics (signatures of shock waves) of 
unprecedented scale (linear separation $\sim$4.4 Mpc at redshift 0.39), 
located at distances from the cluster center which are about 0.7 and 1.3
of the cluster's virial radius.
Another possible giant relic and a radio-halo is detected closer to the 
cluster center. These relic sources are unique ``signposts'' of extremely energetic 
mergers and shocks (both ongoing and past), that are assembling and heating 
up this very massive galaxy cluster. They are also a probe of the filamentary 
cosmic-web structure beyond the cluster virial radius. Optical imaging 
with the IUCAA 2m telescope and \xmm\ X-ray data confirm a very rich 
galaxy cluster with a morphologically disturbed core region, suggesting a
dynamically perturbed merging system.

\end{abstract}


\keywords{cosmology: observations -- galaxies: clusters: individual(PLCK G287.0+32.9) -- shock waves -- X-rays: galaxies: clusters}

\section{Introduction}

Two main baryonic  constituents of the diffuse intracluster medium (ICM) 
are the diffuse hot ($T\sim 10^{7} - 10^{8}$ K), tenuous ($n_{0}\sim$ 10$^{-(3 - 4)}$ 
cm$^{-3}$) ICM and a rarely visible non-thermal 
population of extremely high energy relativistic particles. The presence 
of GeV electrons/positrons (of largely unknown origin) is inferred from
the diffuse synchrotron radio emission in a fraction of galaxy clusters,
which lacks an obvious association with any of the cluster galaxies.
These radio sources, which usually possess large size ($\sim$0.1 - 1 Mpc)
and steep spectra ($\alpha \sim$ 1 - 3; S$_{\nu}$ $\propto$ $\nu^{-\alpha}$),
are called {\em ``radio halos''}\, if they permeate the cluster centers and
{\em ``radio relics''}\, if they are located at the cluster peripheral regions 
(e.g., the review by Ferrari et al. 2008).

Since radio halos and relics are exclusively found in the clusters
clearly showing X-ray or optical substructure \citep{Govoni04}, their 
origin is related to the merging processes of galaxy clusters, although
its details are still obscure. 
Signatures of giant shock waves arising from cluster mergers have been 
discovered in recent years \citep{Bagchi06,vanweeren10}, and such
collisionless shocks are capable of accelerating the particles to 
relativistic energies, giving rise to the 
radio relics via in-situ diffusive shock acceleration of electrons (primary 
electrons). The radio halos may arise via turbulent energy injection or 
in-situ acceleration of protons with the secondary production of relativistic 
electrons and positrons by inelastic $p - p$ collisions (secondary electrons)
\citep{Dennison80,BL07}. 

The free thermal electrons of hot ICM  inverse Compton scatter the photons
of cosmic microwave background (CMBR). This results in a predictable
spectral distortion of CMBR toward the cluster which is the well-known 
thermal Sunyaev-Zel'dovich (SZ) effect \citep{SZ72,SZ80}. The SZ effect is 
an exceptional tool for the study of clusters as the surface brightness of 
SZ signal is independent of redshift and depends only on the integrated 
thermal pressure of ICM along the line-of-sight. The total SZ signal, being 
a measure of thermal energy of the cluster, tightly correlates with the 
cluster mass, making it a powerful, unbiased  tool for making volume limited 
samples and for  finding the high redshift and the most massive clusters 
\citep{Vanderlinde2010,Williamson11}.  Of particular interest is to 
investigate how the SZ signal correlates with other cluster properties, 
such as the mass, temperature and X-ray luminosity (the ``scaling relations''), 
while departure  from scaling  relations indicate non-gravitational and 
non-thermal heating processes in strongly merging systems.

The   \planck\footnote{\planck\ (\url{http://www.esa.int/Planck}) a project
 of the European Space Agency (ESA) with instruments provided by two scientific
consortia funded by ESA member states (in particular the lead countries France 
and Italy), with contributions from NASA (USA) and telescope reflectors 
provided by a collaboration between ESA and a scientific consortium led and 
funded by Denmark.} space telescope designed for the study of CMBR and its 
anisotropy, has recently  also been used for studying the SZ
decrement \citep{planck2011-1.1}. 
The ongoing \planck\ SZ survey provides sky images
at nine frequencies in the range 30--857 GHz and is the first 
all-sky survey capable of blindly searching for distant and massive clusters, 
since the {\em ROSAT} all-Sky Survey (RASS) in the X-ray domain. 

A sample of SZ detected clusters in 
the \planck\  Early Release Catalog  (ESZ)
is described in detail in Planck Collaboration (2011b).  The  highest signal-to-noise 
ratio (S/N $ > 6$)  sample of 189  clusters is
derived from  a blind multi-frequency  search in the all-sky maps from  
the first 10 months of observations. Out of these, 20 are new,
hitherto unknown clusters which are above 
the ESZ high signal-to-noise selection criterion. 
Twenty-five clusters have been observed to date with \xmm\ in snapshot 
exposures ($\sim$ 10 - 20 ks), with 10 in the lower S/N ratio 
category ($ 4 <$ S/N $ < 6$) and 15 in high-significance range with S/N $ > 5$. 
A total of 21 cluster candidates are validated as extended X-ray sources, 
the majority of which show highly irregular and dynamically perturbed structure 
of merging systems  \citep{planck2011-9.1}. 

In this Letter we report the discovery of giant ($>$1~Mpc) non-thermal 
peripheral double radio relics of an unprecedented scale (separation $\sim$4.4 Mpc
at redshift 0.39) in a  massive galaxy cluster \plck, detected solely through 
a blind SZ search by \planck. In view of its extreme X-ray properties and 
non-relaxed morphology, we have imaged it with the Giant Metrewave Radio 
Telescope (GMRT), and also obtained deep optical images with the  2m 
telescope at the IUCAA Girawali Observatory (IGO). These observations 
have revealed this cluster to be  highly unusual in terms of its  morphology, 
mass, richness and non-thermal emission,  relating to its complex dynamical 
state.  

We adopt a $\Lambda$CDM cosmology with $H_0=71$~km~s$^{-1}$~Mpc$^{-1}$, 
$\Omega_M=0.27$ and $\Omega_\Lambda=0.73$, resulting in scale of 5.26 
kpc~arcsec$^{-1}$ (or 3.17 arcmin~Mpc$^{-1}$) for a redshift 0.39. Below
$M_{n}$ and $R_{n}$ denote the total mass and radius corresponding to 
a total density contrast $n \rho(z)$, where $\rho(z)$ is the critical density 
of the universe at redshift $z$. 

\section{\plck: An Exceptionally Massive and Hot  \planck\ ESZ Cluster}

Our target \plck\ is the second most significant detection (S/N$=$10.2) 
among the 20 new clusters of the \planck\ ESZ sample 
\citep{planck2011-8.1}.
A 10~ks X-ray validation exposure taken with \xmm\ reveals 
an extremely  hot ($T_{x} = 12.86$ keV), luminous ($L_{(0.1 - 2.4}$ keV) = $1.72 \times 10^{45}$ 
erg\ s$^{-1}$ within $R_{500}$), and massive (estimated $M_{500} = 
1.57 \times 10^{15}$ $M_{\odot}$) galaxy cluster at redshift of $z$ = 0.39 
(from Fe~K line fitting). 
\plck\ is also the most massive and the hottest new SZ cluster detected by 
\planck. 
Our observations with the 2mt telescope (see below) provide the first 
deep optical image, confirming it to be a very rich galaxy cluster.
The X-ray and SZ properties 
are summarised in Table~\ref{table1}, taken from Planck Collaboration (2011c). 
We have reanalyzed the public domain \xmm\ data providing clear 
evidence for significant morphological disturbances in the center, 
thus characterizing \plck\ as a dynamically perturbed, merging system 
(Figure~\ref{fig_1}). 

\section{Observations}

\subsection{Optical Validation  of Galaxy Cluster \plck}

We observed   \plck\  with the  2m optical telescope of the IUCAA 
Girawali Observatory (IGO) in India,  using the IUCAA {\em Faint 
Object Spectrograph \& Camera (IFOSC)}. IFOSC employs an  EEV 2K~x~2K CCD  
giving a field of view of about 10.5~x~10.5 arcminutes.  
Observations were taken on 2011 February 
3rd, 6th, 7th and 10th, in fair seeing (average $\sim$1\arcsec) and dark sky 
conditions. The analysis 
was done in a standard way using IRAF. The final, combined  deep R-filter image 
has  a total exposure time of 4hr (12$\times$1200 s), which shows a
dense concentration of galaxies near the cluster center. Lacking a good  
magnitude calibration of the CCD images, we used  the SuperCOSMOS Sky Survey for
photometric information \cite{Hambly01}. We obtained a plot of galaxy counts by
including galaxies with ESO R-magnitude $m_{R}<$20.5 and counting in cells of 
100\arcsec\ radius. At the centre a peak galaxy over density factor of $\sim$10 was found,
relative to the mean field galaxy density far away. Thus, we 
independently confirm \plck\ to be a very rich and 
massive galaxy concentration (see Figure~\ref{fig_1}). 

\subsection{GMRT and VLA Radio Images}

A search for diffuse non-thermal radio emission  in and around the  
new \planck\ ESZ clusters was carried out, using the TIFR-GMRT Sky Survey (TGSS) database at 150 MHz.
A central radio-halo/relic source and a peripheral giant double-relic were
thus identified towards \plck.
These are corroborated by the 1.4 GHz NRAO VLA Sky Survey images  
\citep[NVSS;][]{nvss}, as described below.

The 150 MHz image presented here was derived from the data of
the ongoing TGSS \footnote{For TGSS survey details, online data and 
analysis  please refer to http://tgss.ncra.tifr.res.in} (S.K. Sirothia et al., 
in preparation). 
We reanalyzed the TGSS snapshots 
with ROBUST = 5 weights (AIPS++ software) in order to enhance the sensitivity 
to extended emission. This map with a beam of 36.84\arcsec\ x 28.58\arcsec\ 
FWHM in 47.6\degrees position angle, has an rms noise of 3--6  mJy beam$^{-1}$ 
(see Fig.~\ref{fig_2}). The 1.4 GHz map in Figure~\ref{fig_3} is reproduced from the NVSS.


\section{Results}

\begin{figure*}
\includegraphics[angle=0,scale=0.5]{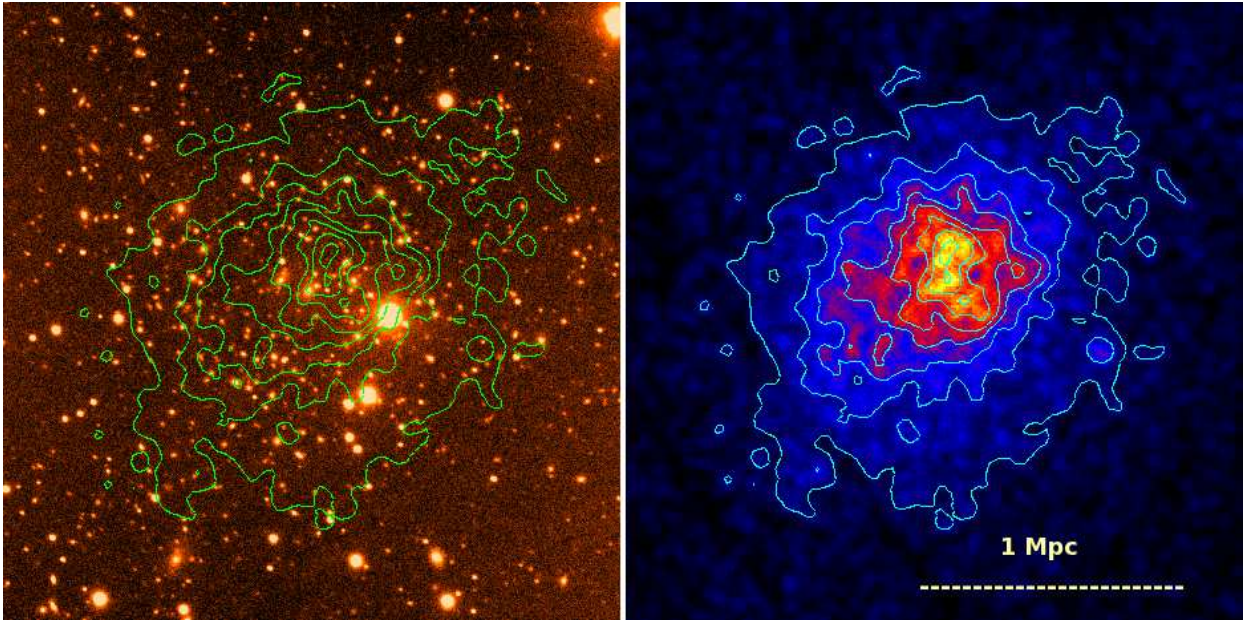}
\caption{Left: deep $R$-band image  taken with  the 2m telescope of IUCAA.
 The FOV is  7 arcmin  on each side ($\sim$2.2 Mpc).  X-ray contours are 
 from \xmm\ (0.3 - 10 keV,  MOS1/2, and PN  detectors, binned and smoothed). Right: 
 the same X-ray image is shown  along with iso-contours.}
\label{fig_1}
\end{figure*}

\begin{figure*}
\includegraphics[scale=0.9]{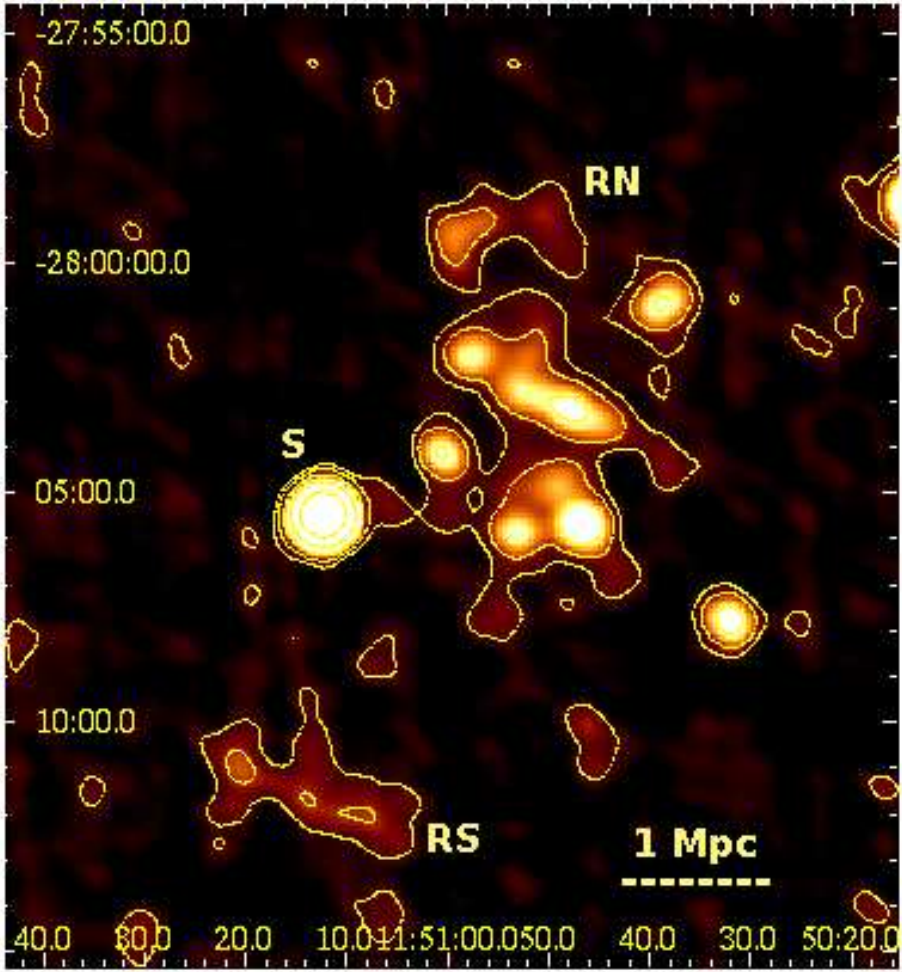}
\caption{Radio image from VLA 1.4 GHz NVSS at 45\arcsec\ FWHM resolution. 
Contours are logarithmically spaced between 0.9 to 60.0  mJy beam$^{-1}$ and
the noise rms is $\sim$0.45  mJy beam$^{-1}$. The pair of peripheral radio relics are marked RN and RS.
Compact source ``S'' is  discussed in the main text.}
\label{fig_3}
\end{figure*}

Our key result is the
discovery of a pair of giant ($l>$1~Mpc), non-thermal peripheral radio 
relics of an unprecedented scale (projected separation $\sim$4.4 
Mpc at z=0.39) in a merging galaxy cluster of extreme properties 
(Figure~\ref{fig_2} top row). Such Mpc-scale double-relics are rare and
only nine cases have been reported.
Of these, the double radio relic associated with the well known 
merging cluster Abell~3667 (z = 0.0556) and having an overall size of
3.9~Mpc is the largest known till now \citep{Rottgering97}.
Furthermore, the present case (\plck) is not only
the first clear detection of diffuse non-thermal radio emission in a galaxy
cluster detected solely through a blind SZ search by \planck, it is also 
the most distant ($z$ = 0.39) giant double radio relic known, even surpassing the
hitherto known case of the
$z=0.27$ filamentary cluster ZwCl~2341.1+0000 \cite{vanWeeren09}. 
In \plck\ the pair of Mpc-scale, sharp ``arc'' like relics are seen to the
north and south of the cluster,
capping the opposite ends of the X-ray emitting hot ICM  and the central optical 
galaxy concentration (Figures~\ref{fig_1} and \ref{fig_2} top row). 
The  two extended radio-relics are marked RN (north relic) and RS (south relic), also shown 
with iso-contours, superposed on the optical images (lower left 
and right panels in Figure~\ref{fig_2}).
Another Mpc-scale Y-shaped filamentary radio feature 
is visible closer to the north of the cluster center,  whose nature is 
presently unclear ($\sim$R.A. $11^{h} 50^{m} 51.30^{s}$, 
decl. $-28^{d} 02^{m} 57.4^{s}$), but which may well be another radio
relic, as it has no obvious association with any galaxy.
Both this feature and  
the peripheral double relic are visible also in the lower resolution 
NVSS image at 1.4 GHz (Figure.~\ref{fig_3}), lending 
independent support to their being real. Moreover, NVSS 
image shows some diffuse radio emission near the cluster center, probably representing
a cluster wide
radio-halo (Figure~\ref{fig_3}). Unfortunately, the poor resolution
and the presence of several discrete radio sources in that region makes
it difficult to unambiguously visualise the proposed halo.

Note that the strong point source marked ``S'' on NVSS image (181 mJy peak) 
lacks a TGSS counterpart, implying a strongly inverted spectrum below
$\sim$1 GHz. On the other hand, the nature of the 
extended Mpc-scale feature marked `F' remains unclear; although seen in the 
TGSS 150 MHz image (Figure~\ref{fig_2}), it is absent in the NVSS 1.4 GHz map. 
This suggests that  either it is a very steep spectrum radio-relic, possibly
a southern counterpart to the inner relic, or merely a noise feature.
To clarify this we have applied for much deeper GMRT imaging.

Integrated flux densities of the radio relics  are; 550 $\pm$ 50 mJy (RN) 
and 780 $\pm$ 50 mJy (RS) at 150 MHz (GMRT), and 33 $\pm$ 5 mJy (RN) and 
25 $\pm$ 5 mJy (RS) at 1400 MHz (Very Large Array, VLA). Thus, the radio spectral index 
$\alpha$ between the two frequencies is extremely steep, i.e., 1.26 for 
the  relic RN and 1.54 for the  relic RS. The corresponding value for the
giant Y-shaped relic candidate north of the cluster center is also very 
steep at $\alpha=$1.20. These very steep spectra support the interpretation
of all these giant diffuse sources as being radio relics, an inference
corroborated by 
the lack of obvious  association with any prominent galaxy (Figure~\ref{fig_2}). 
We also point out that the radio emission from a foreground ($m_{R}$=15.34),  
emission-line galaxy 2MASX~J11504002-2800582 at $z$ = 0.0605 \citep{MS07} 
is seen superposed on the western end of the relic RN. 


Our reanalysis of the 10~ks exposure \xmm\ archival data for \plck\ shows an 
extremely X-ray luminous cluster having prominent 
substructure within the central 1.5\arcmin ($\sim$500 kpc) core region 
(Figure~\ref{fig_1}). The brightest central elliptical galaxy (ESO R magnitude 17.56) 
at R.A. $11^{h} 50^{m} 50.1^{s}$, Dec. $-28^{d} 04^{m} 55.3^{s}$ is located 
within a subcluster, which is clearly shifted to the south-east from the 
main X-ray peak by $\sim$1.3\arcmin\ (400 kpc). 
The compression of the X-ray contours north-west of the main X-ray peak 
indicates ongoing gas motions, either due to gas ``sloshing'' following a merger, or motion 
of a merging subcluster unit. A detailed analysis of the X-ray data will be presented in
J. Bagchi et al. (in preparation).


\section{Discussion and Conclusions}

The giant radio relics reported here in the cluster \plck\ are not only 
the first such detections for an SZ cluster survey, but
also this relic pair has the largest separation (4.4 Mpc) and redshift
($z$ = 0.39) compared to any other giant double relic known.
In \plck\ the projected distances from the X-ray center of 
the cluster are about 5\arcmin\ (or 1.58 Mpc) for the northern relic and 
9.5\arcmin\ (or 3 Mpc) for the southern relic. The virial radius ($R_{vir}$) 
of cluster  can be estimated by taking $R_{vir}\approx$$R_{200}$ and  
scaling from $R_{500}$, i.e., 
$R_{200}$ = $R_{500}$$\times (500/200)^{1/3}$$\times (M_{200}/M_{500})^{1/3}$.
This gives $R_{200}$$\approx$2.3 Mpc, using $R_{500}$ = 1541 kpc and the mean 
value of mass ratio ($M_{200}/M_{500})=$1.40$\pm$0.02, obtained from the Navarro-Frenk-White (NFW)
profile fit to nearby ($z$\lapp0.15) clusters \citep{Pointe05}. 
Thus the separation from the cluster centre of the relics is
about 1.3$\times$$R_{200}$ for RS and
about 0.7$\times$$R_{200}$ for RN.
The mechanism of electron acceleration at such extreme distances from the
cluster center is little understood. According to a currently popular 
interpretation, double-relics 
are the sites where cosmic-ray particles are accelerated in-situ to 
relativistic energies by diffusive first-order Fermi acceleration, such
that the resulting radio emission traces the outward expanding powerful
shock fronts originating from a major cluster merger activity at the 
center \cite{Ensslin98,Bagchi06,Hoeft_et_al._2008,vanweeren10}. 

\begin{figure*}
\includegraphics[scale=0.55]{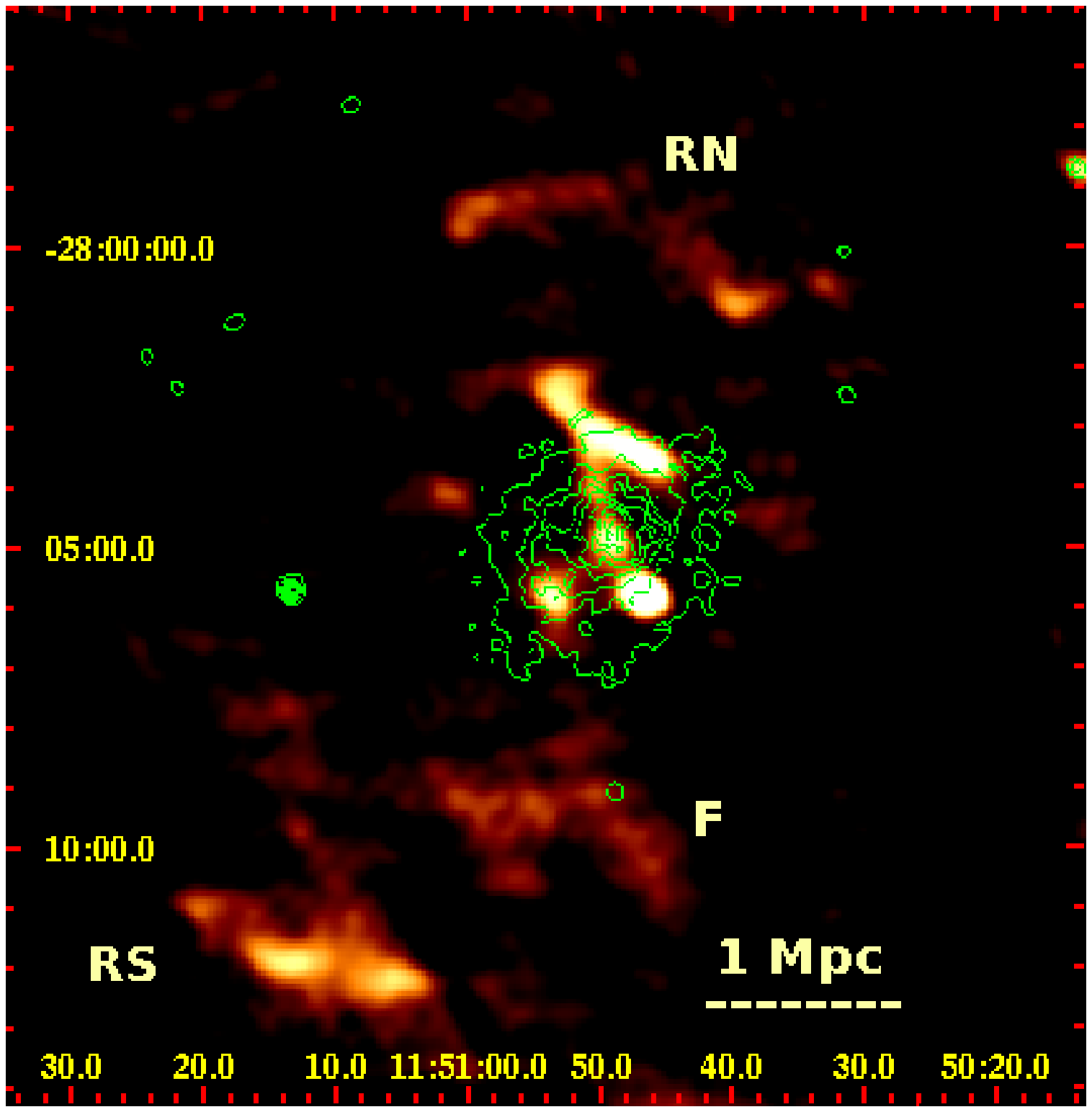}
\includegraphics[scale=0.5]{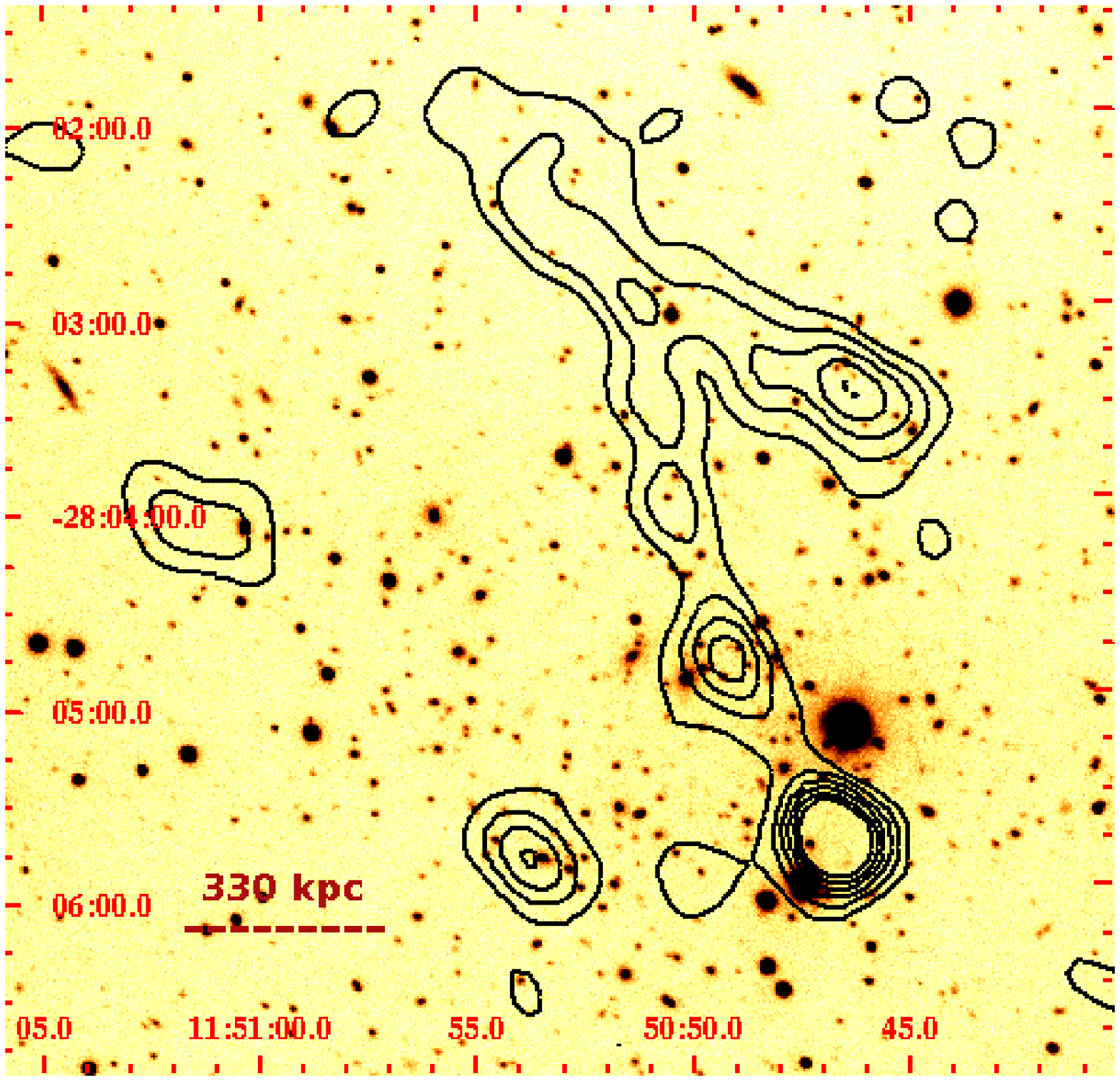}
\includegraphics[scale=0.374]{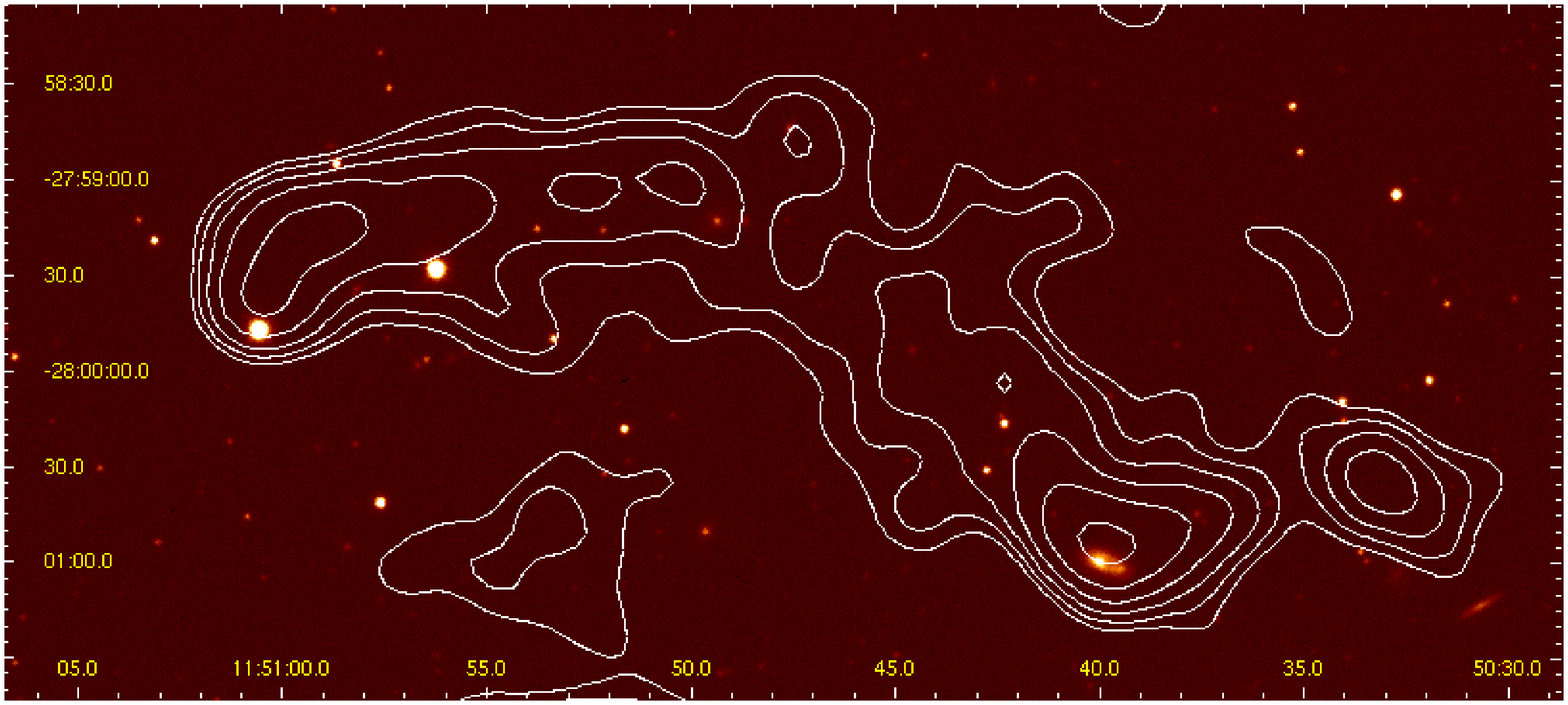}
\includegraphics[scale=0.33]{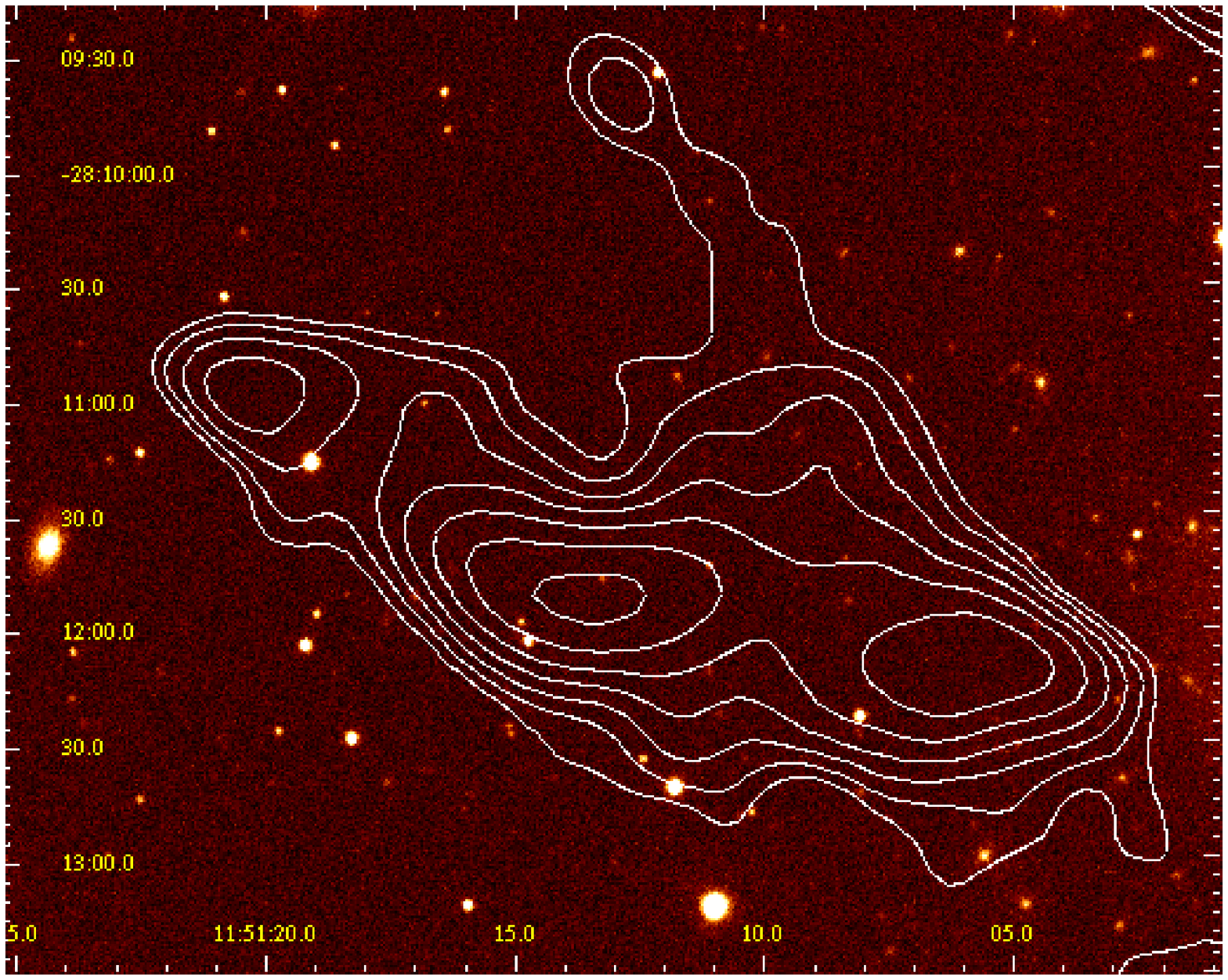}
\caption{Upper left: GMRT 150 MHz radio map of the cluster (orange-yellow, beam FWHM
36.84\arcsec x 28.58\arcsec in 47.6\degrees\ P.A.). Superposed are the \xmm\,
X-ray emission contours (green) in 0.3 - 10 keV energy range. A pair of giant radio relics 
at the cluster periphery are marked RN (north relic) and RS (south relic). Upper right:
GMRT  150 MHz radio map near the cluster center 
(beam FWHM 24\arcsec x 15\arcsec\ in P.A. 30\degrees), superposed on the optical  $R$-band image.
Contour levels are: 7.5, 13.57, 19.64, 25.7, 31.75, 43.92,  and 50 mJy beam$^{-1}$.
Lower left and right: Contour plots of radio relics RN (left) and RS (right)  
at 150 MHz, superposed on the  R-band image of the region. 
The beam FWHM is 36.84\arcsec x 28.58\arcsec in 47.6\degrees\ P.A.. 
Contour levels are: 10, 13, 17, 22, 29, 33, and 50
 mJy beam$^{-1}$  with typical rms noise $\sim$3--5 mJy beam$^{-1}$.}
\label{fig_2}
\end{figure*}

This scenario is strongly supported by the extremely high temperature and 
unrelaxed X-ray appearance of \plck.
A recent  hydro-simulation of major mergers \citep{Paul2011} show that
the emergent ellipsoidal shock front may expand well beyond the virial 
radius and the brightest parts of it are roughly tangential 
to the principal merger axis, thus resembling  a pair of  concave ``arcs'', akin to those
found in \plck. As the shock front crosses the virial
radius, it begins to interact with the intergalactic matter filaments,
resulting in its fragmentation into segments.
Such predicted evolution of the peripheral shocks appears to be mirrored
in both members of the relic pair, whose overall arc-shaped radio morphology 
exhibits one or two distinct emission gaps (Figure~\ref{fig_2}).
Thus, such giant radio relics can be a new probe of the large-scale 
cosmic structures \citep[see also][]{Bagchi06,vanweeren10,Brown11}.

\xmm\  data confirm that the cluster \plck\ is exceptional in 
terms of its mass, temperature, and luminosity  (Table~\ref{table1}),  
detected with high S/N by \planck\ \cite{planck2011-9.1}. 
Given the  perturbed dynamical state of merging system \plck, it is of
interest to investigate its position on the well-known X-ray--SZ scaling planes. 
As shown in \cite{planck2011-9.1}, indeed \plck\ is a prominent outlier
from both $L_{x,500}$--$Y_{500}$ and $M_{500}$--$L_{x,500}$ mean scaling lines, which are obtained
from the bias corrected local REXCESS X-ray sample \citep{arnaud10}, and a sample of 62
nearby galaxy clusters detected by \planck\, showing  strong SZ decrement \cite{planck2011-11.1}.   
Although due to the intrinsic scatter in these correlations, the deviation of \plck\ from
these scaling laws is not more than 1$\sigma$, nevertheless it is noticeably underluminous 
in X-rays for its large mass, in common with other new extreme low luminosity, high mass, disturbed 
clusters that are being revealed by \planck.  


Synchrotron emission from the giant relic pair shows that magnetic fields 
of appreciable strength are present not only in the ICM but also in the 
diffuse intergalactic medium beyond, i.e., in the gas that will be shocked 
as it accretes onto collapsing structures -- the precursors of virialized 
galaxy clusters. Such magnetic fields are also required for providing the 
scattering centers for the diffusive shock acceleration mechanism.
As it is not obvious how magnetic fields are amplified up to the requisite large 
values along the filaments, studies similar to the one reported here
pose a new challenge to theoretical models. The detection of large scale 
diffusive shocks via synchrotron radio emission in the sparsely studied, 
rarefied intergalactic 
field around the cluster-outskirts, i.e., beyond the virial radius 
($R\gtrsim$ $R_{200}$), is a very significant result, and it provides a 
foretaste of science with the 
upcoming low frequency radio telescopes like LOFAR and LWA (and SKA at 
higher frequency), as an effective probe of the non-thermal processes in 
merging clusters and within the filamentary cosmic-web of the near and 
distant parts of the universe (e.g. Rudnick et al. 2009).

\acknowledgments
We thank the staff of the GMRT and IUCAA Girawali Observatory 
that made these observations possible. We thank Monique Arnaud and Gabriel Pratt
for useful discussions. GMRT is run by the 
National Centre for Radio Astrophysics (NCRA) of the Tata Institute of Fundamental Research (TIFR).
The X-ray data are from {\em XMM-Newton}, a European Space Agency (ESA) 
science mission with instruments and contributions directly funded by ESA Member States and NASA.
MBP was supported by the ``INSPIRE Fellowship'' program of 
DST, Ministry of Science and Technology, Government of India. 
N.W. was supported by the NASA through Chandra Postdoctoral 
Fellowship Award Number PF8-90056 issued by the Chandra X-ray Observatory Center, which is 
operated by the Smithsonian Astrophysical Observatory for and on behalf of the 
National Aeronautics and Space Administration under contract NAS8-03060.


\newpage

\begin{table}
\begin{minipage}{85mm}
\caption{\label{table1}
}
\begin{tabular}{@{}l@{}l@{}}
 {\large X-ray and SZ Parameters}\footnote{Planck Collaboration (2011b, 2011c)}\\
 \hline
 Right Ascension &  $11^{h} 50^{m} 49.20^{s}$\\
 Declination & $-28^{d} 04^{m} 36.5^{s}$\\
 Redshift\footnote{From fitting Fe-K lines} & 0.39\\
 X-ray temperature  &  $12.86 \pm 0.42$ keV\\
 X-ray luminosity\footnote{Within $R_{500}$, in (0.1 - 2.4\ keV) band} & $17.20 (\pm 0.11) \times 10^{44}$ ergs$^{-1}$\\
 Cluster mass & $15.72 (\pm 0.27) \times 10^{14}$ $M_{\odot}$\\
 $R_{500}$ & 1541 kpc\\
 SZ Compton $Y_{500}$\footnote{Spherically integrated Compton $Y$-parameter measured
 interior to $R_{500}$} & $ 3.30 (\pm 0.16) \times 10^{-4}$ Mpc$^{2}$\\
 X-ray $Y_{X,500}$\footnote{X-ray derived $Y$-parameter. $Y_{X,500}$ = $M_{g,500} T_{x}$, the product
 of gas mass within $R_{500}$ and X-ray temperature} & $ 30.69 (\pm 0.36) \times 10^{14}$ $M_{\odot}$ keV\\
 \hline
 \end{tabular}
 \end{minipage}
 \end{table}


\end{document}